\documentclass[journal]{IEEEtran}

\usepackage{graphicx}
\usepackage{fancyhdr}
\fancypagestyle{firststyle}

\begin{document}
%
\title{PETALO: Time-of-Flight PET with liquid xenon}



%

\author{Paola~Ferrario, Vicente~Herrero-Bosch, Jos\'e~Mar\'ia~Benlloch-Rodr\'iguez, Carmen~Romo-Luque \\and~Juan~Jos\'e~G\'omez-Cadenas
\thanks{P. Ferrario is with Donostia International Physics Center, 
Paseo Manuel de Lardizabal 4, 
20018 Donostia-San Sebasti\'an, Spain, e-mail: paola.ferrario@dipc.org.}
\thanks{V. Herrero-Bosch is with Instituto de Instrumentaci\'on para Imagen Molecular, CSIC-Universitat Polit\`ecnica de Val\`encia, Camino de Vera s/n,  46022 Valencia, Spain.}
\thanks{Jos\'e Mar\'ia Benlloch-Rodr\'iguez and Carmen Romo-Luque are with Instituto de F\'isica Corpuscular,  CSIC-Universitat de Val\`encia, Calle Catedr\'atico Jos\'e Beltr\'an, 2, 46980 Paterna, Spain.}
\thanks{Juan Jos\'e G\'omez-Cadenas is with Donostia International Physics Center,  Paseo Manuel de Lardizabal 4, 20018 Donostia-San Sebasti\'an, Spain.}
}


%


\maketitle

\thispagestyle{firststyle}
\renewcommand{\headrulewidth}{0pt}

\begin{abstract}
The fast scintillation decay time and the high scintillation yield of liquid xenon makes it an appropriate material for nuclear medicine. 
Moreover, being a continuous medium with a uniform response, liquid xenon allows one to avoid most of the geometrical distortions of conventional detectors based on scintillating crystals. In this paper, we describe how these properties have motivated the development of a novel concept for positron emission tomography scanners with Time-Of-Flight measurement, which uses liquid xenon as a scintillating material and silicon photomultipliers as sensors. Monte Carlo studies have indicated that this technology would provide a very good intrinsic time resolution, of around 70 ps. Moreover, being liquid xenon transparent to UV and blue wavelengths, both scintillation and Cherenkov light can be exploited. While the former can be used for energy measurements, the latter is a prompt signal (of a few picoseconds), which provides a very precise time measurement.
Monte Carlo simulations point to a time resolution of 30-50 ps obtained using Cherenkov light. A first prototype is being built to demonstrate the high energy, spatial and time resolution of this concept, using a ring of around 30 cm of internal diameter and a depth of 3 cm instrumented with VUV--sensitive silicon photomultipliers.
\end{abstract}


%
\IEEEpeerreviewmaketitle

\section{Introduction: liquid xenon in PET imaging}
\label{sec:lxe}

Liquid xenon (LXe) is a scintillator material widely used in particle physics, especially in rare event searches such as dark matter or neutrinoless double beta decay. When a high energy gamma of 511 keV interacts with LXe, a photoelectric or a Compton interaction can occur, producing an electron which propagates in the medium, releasing its energy via scintillation (with an average wavelength of 178 nm) and ionization. 
The LXe density is reasonably high (2.98 g/cm$^3$), and provides an attenuation length to 511-keV photons of 3.7 cm. Together with an acceptable Rayleigh scattering length of 36 cm, this makes LXe a suitable material for positron emission tomography (PET) scanners. LXe presents a high scintillation yield  ($\sim$ 30\,000 photons per 511 keV gamma, without applied electric fields) and a fast scintillation decay time (2.2 ns in its fastest mode), which are essential for Time-of-Flight (TOF) applications. 
Also, LXe is transparent to its own Cherenkov light, which opens the possibility of an ultra fast coincidence detection, achieving, therefore, a very high precision in the coincidence resolving time (CRT).

A liquid scintillator medium is continuous and provides a uniform response, which makes it feasible to design a compact system even with large dimensions. A high precision 3D measurement of the position of the gamma interactions is possible, including the depth of interaction (DOI). Furthermore, xenon liquefies at a temperature of 160 K, at atmospheric pressure, therefore he system can be operated with a simple cryostat.

The first one who proposed to use LXe as a scintillator to build a PET was Lavoie in 1976 \cite{lavoie}. In the 1990's Chepel and co-workers  built a Time Projection Chamber (TPC) and read both the scintillation and ionization signal \cite{chepelFirst}, while Doke and Masuda proposed to read only scintillation light with photomultipliers  \cite{DokeMasuda}.
In 2007, the Xemis group at Subatech started building a LXe Compton telescope, which uses a $\beta^+$ decay isotope that also emits a high energy gamma, such as $^{\ensuremath{44}}$Sc \cite{GallegoManzano:2015hkg}. 


These studies didn't exploit all the potential of LXe applied to PET. On one hand, although reading both light and charge gives a better energy and spatial resolution, a TPC is slow and introduces dead time due to charge drift, which may decrease the sensitivity, and increases the complexity and cost of the detector, thus jeopardizing the large scale implementation of the technology. On the other hand, the large size of PMTs results in less precise geometric corrections and a worse spatial resolution. 

\section{The PETALO approach}

A new approach has been proposed recently \cite{Gomez-Cadenas:2016mkq, Gomez-Cadenas:2017bfq}, based on liquid xenon scintillation. The key idea of PETALO (a Positron Emission Tof Apparatus based on Liquid xenOn) is to obtain good energy, position and time resolution, collecting the light produced by xenon scintillation.
To this end, the sensors chosen for the light readout are SiPMs, which provide large area, high gain and very low noise. At the liquefaction temperature of xenon, SiPMs show a negligible dark count rate and a perfect functioning.

This approach overcomes the weak points of previous attempts to use LXe in PET technology, since no high voltages and dead times due to drift are needed. Moreover, it exploits new generation sensors, which are smaller, therefore can be used to instrument also the entry face of the detector, thus providing a better spatial resolution in the reconstruction of the gamma interaction.

\section{Monte Carlo studies and results}

\begin{figure}[htbp]
\centering
\includegraphics[width=0.4\textwidth]{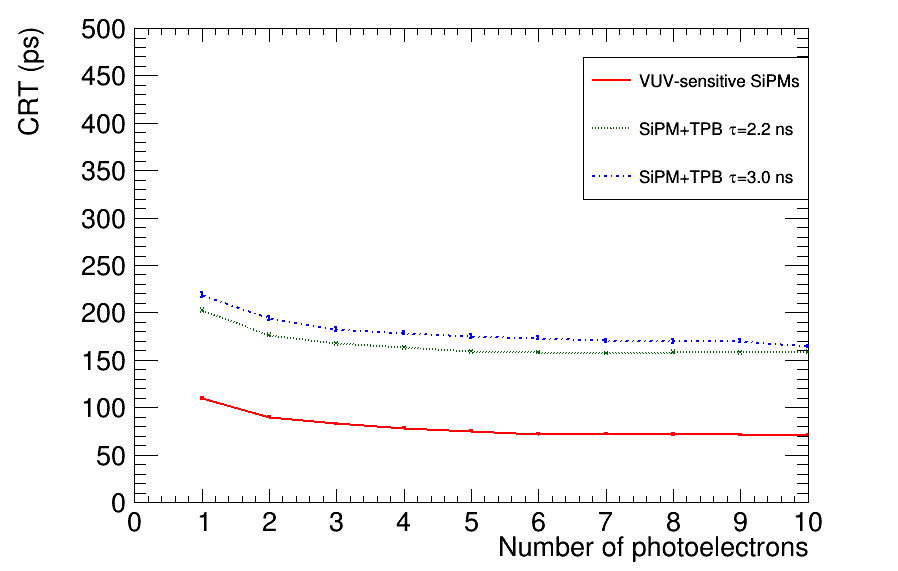}
\caption{\label{fig:tof} Coincidence resolving time as a function of the number of photoelectrons used for the calculation. Figure from \cite{Gomez-Cadenas:2016mkq}.}
\end{figure}

A Monte Carlo study has been performed, to assess the potential of this approach, assuming a setup made of two opposite cubic cells, where each cell has the entry and the exit faces instrumented, while the others are covered with a highly reflective material such as PTFE, which features close to $100\%$ reflective efficiency at UV wavelenghts. This study has shown that a very good CRT of down to 70 ps FWHM can be reached using VUV sensitive SiPMs with a photodetection efficiency of $20\%$ \cite{Gomez-Cadenas:2016mkq}, as illustrate in Fig.~\ref{fig:tof}.
%

One of the most attractive features of liquid xenon is the possibility of building a large, uniform volume of scintillation material, which is easier to fabricate, can be purified continuously and has a homogeneous  response, thanks to greatly reduced border effects. 
\begin{figure}[htbp]
\centering
\includegraphics[width=0.5\textwidth]{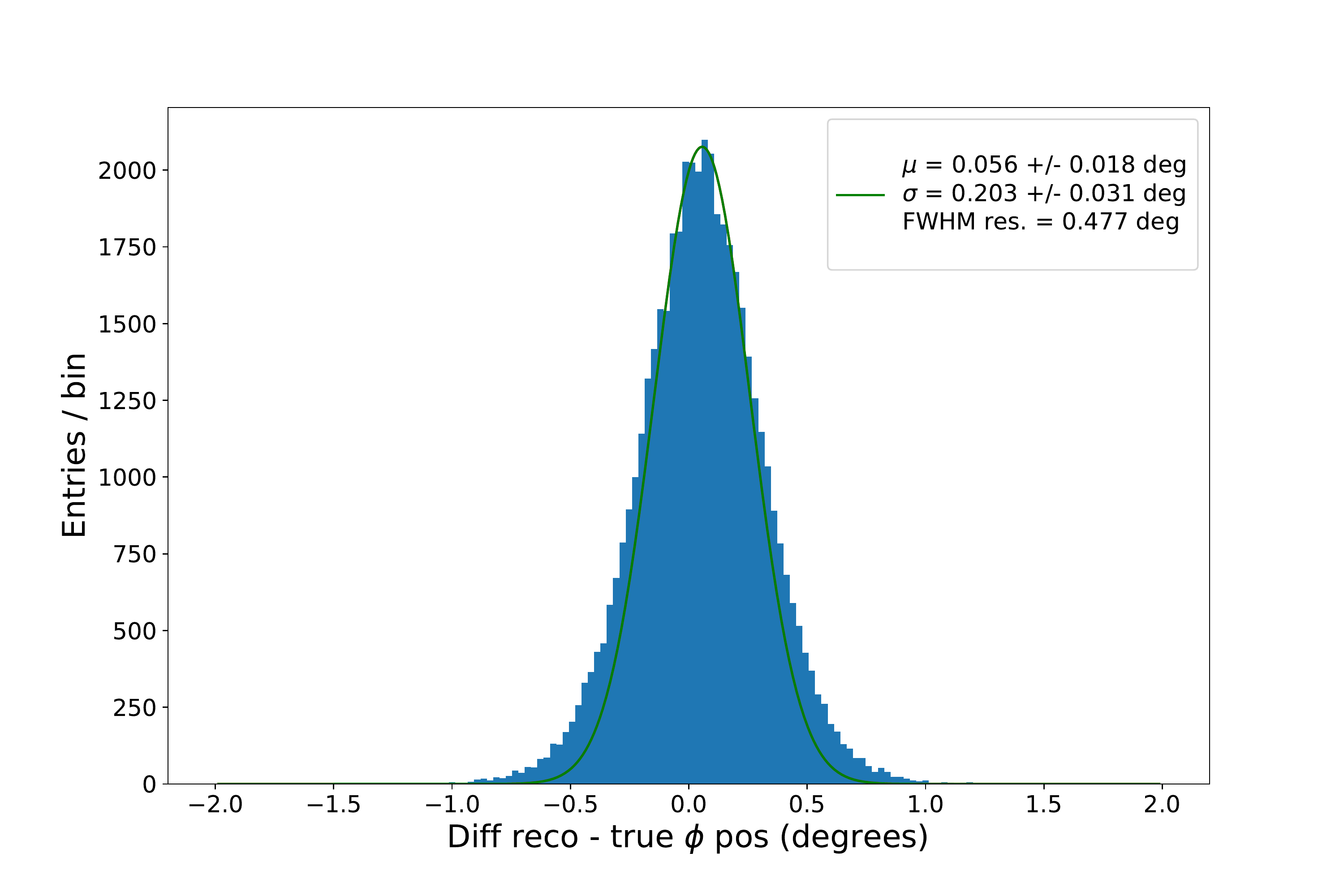}
\caption{\label{fig:phires} Resolution in the determination of the angular position of photoelectric events in a LXe ring.}
\end{figure}
The simulation of a small ring of 165 mm internal radius and 3 cm of depth of interaction provides excellent angular (1.2-1.5 mm FWHM in the arc, shown in Fig. \ref{fig:phires}) and axial (1.6 mm FWHM) resolutions in the interaction position of the gamma, for central DOI and photoelectric events.  DOI resolution in the peak is below 2 mm FWHM. Deviations from gaussian shape are under study.
The energy resolution is dominated by the intrinsic fluctuations of scintillation light in LXe and is estimated to be $\sim20\%$ FWHM.
The intrinsic CRT (without including electronics jitter) is less than 20 ps, therefore we can conclude that the measurement of time in liquid xenon will be dominated by the fluctuations introduced by the electronics. 

A very interesting aspect of the PETALO concept is the possibility of pursuing a Cherenkov-based PET. Using Cherenkov light for TOF is very attractive, due to the promptness of these photons, but has the drawback of a low yield and a high absorption rate in conventional crystal detectors. The distribution of Cherenkov light in terms of wavelengths follows a $1/\lambda^2$ trend, thus most of Cherenkov light is produced in the UV range. Typical crystals such as LSO have a drop in transparency for wavelengths lower than 400 nm, thus most of Cherenkov light is not detected. 
On the contrary, xenon has the advantage of being transparent to UV and blue light, which increases light collection and, therefore, the rate of detection of coincidences. A study has been published \cite{Gomez-Cadenas:2017bfq} which shows that a CRT of 30--50 ps can be achieved, using very fast photosensors, sensitive to wavelengths down to $\sim$ 300 nm, such as microchannel plates.

\section{First prototype}

A first prototype, PETit,  is being currently built, consisting of a cryostat hosting a ring of LXe of $\sim$ 16 cm internal and 19 cm external radii. Both the internal and external walls of the ring are instrumented with VUV-sensitive silicon photomultipliers at a fixed pitch. The ring is continuous, with no sections, which has the advantage of producing less border effects in the reconstruction of the interaction of the gammas. The non-instrumented walls are completely absorbent to light, in order to enhance the spatial resolution. The SiPM signals are extracted via PCB feedthroughs and read by a set of ASICs through connectors made of liquid crystal polymer (LCP), resistant to extreme temperatures.
%

\section{Conclusions}

We have presented a proof-of-concept of  an efficient Time-of-Flight PET scanner based on liquid xenon. Simulations show the potential of this technique and have motivated the construction of a first prototype, which will serve as a demonstrator of energy, spatial and time resolution.

\section*{Acknowledgment}

The authors would like to thank the support of the following institutions and agencies: the European Research Council (ERC) under Starting Grant 757829-PETALO; the Spanish Ministry of Economy and Competitiveness under project FPA2016-78595-C3-3-R; the GVA of Spain under grant PROMETEO/2016/120.



\bibliographystyle{IEEEtran}
%



\bibliography{biblio}

\begin{thebibliography}{1}
\providecommand{\url}[1]{#1}
\csname url@samestyle\endcsname
\providecommand{\newblock}{\relax}
\providecommand{\bibinfo}[2]{#2}
\providecommand{\BIBentrySTDinterwordspacing}{\spaceskip=0pt\relax}
\providecommand{\BIBentryALTinterwordstretchfactor}{4}
\providecommand{\BIBentryALTinterwordspacing}{\spaceskip=\fontdimen2\font plus
\BIBentryALTinterwordstretchfactor\fontdimen3\font minus
  \fontdimen4\font\relax}
\providecommand{\BIBforeignlanguage}[2]{{%
\expandafter\ifx\csname l@#1\endcsname\relax
\typeout{** WARNING: IEEEtran.bst: No hyphenation pattern has been}%
\typeout{** loaded for the language `#1'. Using the pattern for}%
\typeout{** the default language instead.}%
\else
\language=\csname l@#1\endcsname
\fi
#2}}
\providecommand{\BIBdecl}{\relax}
\BIBdecl

\bibitem{lavoie}
L.~Lavoie, ``Liquid xenon scintillators for imaging of positron emitters,''
  \emph{Medical Physics 3}, vol.~5, p. 283, 1976.

\bibitem{chepelFirst}
V.~Chepel, ``{A new liquid xenon scintillation detector for positron emission
  tomography},'' \emph{Nucl. Tracks Radiat. Meas.}, vol.~21, pp. 47--51, 1993.

\bibitem{DokeMasuda}
T.~Doke and K.~Masuda, ``{Present status of liquid rare gas scintillation
  detectors and their new application to gamma-ray calorimeters},'' \emph{Nucl.
  Instrum. Meth.}, vol. A420, pp. 62--80, 1999.

\bibitem{GallegoManzano:2015hkg}
L.~Gallego~Manzano \emph{et~al.}, ``{XEMIS: A liquid xenon detector for medical
  imaging},'' \emph{Nucl. Instrum. Meth.}, vol. A787, pp. 89--93, 2015.

\bibitem{Gomez-Cadenas:2016mkq}
J.~J. Gomez-Cadenas \emph{et~al.}, ``{Investigation of the Coincidence
  Resolving Time performance of a PET scanner based on liquid xenon: A Monte
  Carlo study},'' \emph{JINST}, vol.~11, no.~09, p. P09011, 2016.

\bibitem{Gomez-Cadenas:2017bfq}
J.~J. Gomez-Cadenas, J.~M. Benlloch-Rodr\'iguez, and P.~Ferrario, ``{Monte
  Carlo study of the Coincidence Resolving Time of a liquid xenon PET scanner,
  using Cherenkov radiation},'' \emph{JINST}, vol.~12, no.~08, p. P08023, 2017.

\end{thebibliography}

\end{document}